\documentclass[aps,pra,showpacs,preprintnumbers,amsmath,amssymb]
{revtex4}
\usepackage{graphicx}
\usepackage{bm}

\newcommand{\be}{\begin{equation}}
\newcommand{\ba}{\begin{eqnarray}}

\newcommand{\bF}{{\mathbf F}}

\newcommand{\bN}{{\mathbf 0}}

\newcommand{\ee}[1]{\label{#1}\end{equation}}
\newcommand{\ea}[1]{\label{#1}\end{eqnarray}}
\newcommand{\tr}{\operatorname{tr}}

\newcommand{\Tr}{\operatorname{Tr}}

\begin{document}
\title{Is the dynamics of open quantum systems always linear?}
\author{Karen M. \surname{Fonseca Romero}}
\altaffiliation[On leave from ]{Departamento de F$\acute{\mbox{\i}}$sica, 
Universidad Nacional, Bogot$\acute{\mathrm{a}}$, Colombia.}
\author{Peter Talkner}
\author{Peter H\"anggi}
\affiliation{Institut f\"ur Physik, Universit\"at Augsburg, 
Universit\"atsstr.1, D 86315 Augsburg, Germany}
\date{\today}

\begin{abstract}
We study the influence of the preparation of an open quantum system on
its reduced time evolution. In contrast to the frequently considered
case of an initial preparation where the total density matrix
factorizes into a product of a system density matrix and a bath
density matrix  the time evolution generally is no longer governed by
a linear map nor is this map affine. Put differently, the evolution is
truly nonlinear and cannot be cast into the form of a linear map plus
a term that is independent of the initial density matrix of the
open quantum system. As a consequence, the inhomogeneity that emerges
in formally exact generalized master equations is in fact a nonlinear
term that vanishes  for a factorizing initial state. The general
results are elucidated with the example of two interacting spins
prepared at thermal equilibrium with one spin subjected to an external
field. The second spin represents the environment. 
The field allows the preparation of mixed density matrices of
the first spin that can be represented as a convex combination of two
limiting pure states, i.e. the preparable reduced density matrices
make up a convex set. Moreover, the map from these reduced density
matrices onto the corresponding density matrices of the total system  
is affine only for vanishing coupling between the spins. In general, the
set of the accessible total density matrices is nonconvex.

\end{abstract}
\pacs{03.65.Yz, 05.30.Ch, 02.50.-r}
\maketitle

\section{Introduction}

Closed systems are known to be an idealization. 
In general, real systems interact with their 
environment and exhibit properties that cannot be observed in
finite closed systems, such as irreversibility of the time evolution and,
related, the relaxation of observables toward stationary values   
and dephasing, or
decoherence. Various techniques have been developed to 
treat the dynamics of  open systems without explicitly considering 
the full Hamiltonian dynamics \cite{w}. 
For example, effective equations for the reduced
density matrix of the considered open system, known as master equations,
have been proposed long
ago \cite{pauli,vH,Haa} and still are of considerable interest because of their
conceptional simplicity and potential usefulness \cite{B,MK}. 
New challenges in this field of fundamental physics have
come from nanotechnology \cite{nt} and quantum computing \cite{qc}. 
 
Any equation determining the time evolution of
a density matrix  has to obey several general properties 
which guarantee that the density matrix stays selfadjoint, 
positive and normalized in the course of time.  
These general properties still leave much freedom and, in order to further 
restrict possible dynamical laws, additional requirements 
for the dynamics have been postulated \cite{Su}. One seemingly natural
property that
often is assumed without even being mentioned is the {\it linearity} of the
time evolution, which generally is understood as a consequence of the linearity
of the 
Schr\"odinger and the Liouville-von Neumann equation for closed
systems. This argument, however, only works by analogy and no proof of
the necessity of this requirement is available. Just on the contrary  
Pechukas \cite{P,PC} has recently shown
that linearity may only hold if the
initial state of the total system factorizes into a product of a 
density matrix for the open system and another 
one for the environment, and if a sufficient number of pure states can
be prepared. 
The assumption of linearity is a prerequisite  
of a Markovian dynamics and 
of complete positivity \cite{gks,L}. These properties then lead to the
mathematically well characterized class of Lindblad master equations. 
From the physical point of view, however, these equations suffer from certain
deficiencies \cite{P}. They are restricted to the regime of weak coupling
between the considered system and its environment. In particular 
the weak coupling assumption will fail at sufficiently low 
temperatures \cite{gwt,rhw}. 
Moreover, there are
various general statistical mechanical properties that are
incompatible with the assumption of a Markovian dynamics \cite{t86}. 

A few microscopic models of systems interacting with their environment
can be reduced exactly to Lindblad master equations with time
dependent coefficients \cite{kmfr,kmfr2}, thereby describing the single time 
non-Markovian reduced dynamics. 
          
The dynamics of an open system is determined by both, the full dynamics of
the
considered system interacting with its environment and  the 
initial state of the complete system. The significance of the initial
state was emphasized in several works \cite{ght,gsi,gth,gtht,KG,Be,O,CS,HR}. 
In an experiment this 
initial state is imposed by a preparation procedure. Here we will
only be concerned with equilibrium preparations that  lead to a thermal
equilibrium of the total system in the presence of external fields
that only act on the system and that are switched off finally. In this way, 
the initial state is described by a canonical density matrix:  
\be
\rho^\mathbf{F} 
= Z^{-1} \exp \left \{-\beta \left ( H - \sum_j F_j X_j \right ) \right \},
\ee{rb}
where $\beta$ is the inverse temperature, 
$H$ the Hamiltonian governing the dynamics of the total system, 
$\mathbf{F} = (F_j)$ are external, i.e. classical, fields, $X_j$ the
corresponding conjugate operators of the open system and  
$Z = \Tr  \exp \left \{-\beta \left ( H + \sum_j F_j X_j \right ) \right \}$ 
is the partition function of the total system. Here,
$\Tr$ denotes the trace over the total system. In this way, initial
states of the total system are reproducibly prepared.  They correspond
to a thermal equilibrium of the environment at a given temperature and
to a
state of the system that depends on the strengths of the
external fields $F_j$. 
The set of density matrices that can be obtained upon variation of the
fields forms the
equilibrium preparation class. 
The reduced states belonging to this preparation
class are determined by the trace over the environment, which is
denoted by $\Tr_B$:
\be
\rho^\mathbf{F}_S = \Tr_B \rho^\mathbf{F}.
\ee{rS}
The calculation of this trace 
is nontrivial in most cases and in general does not
lead to the canonical distribution of the uncoupled system, $Z^{-1}_S \exp
\left \{ - \beta H_S \right \} $ \cite{gwt,zt,K,I} 
where $H_S$ is the Hamiltonian of the system in
presence of the external fields and $Z_S$ the respective partition
function. This particular form is only
obtained in the limit of weak coupling between the system and the
environment \cite{vH,K,Sp}. In the weak coupling limit, the equilibrium 
density matrix
of the total system factorizes into a product of a system and an
environment 
density matrix. This is an example of the factorizing preparation
which leads to a product of a particular density matrix of the
environment 
$\rho_B$ and an arbitrary density matrix $\rho_S$ of the system: 
\be
\rho^{\mathrm{fac}} = \rho_S \ \rho_B.
\ee{rfac}
The factorizing preparation is assumed in most
theoretical investigations though it is often difficult, if not
impossible, to realize it experimentally.

A more general class of preparations has been suggested  
in the context of the path integral approach to
open systems \cite{gsi}:
\be
\rho^O = \sum_j O_j \rho^\mathbf{F} O_j',
\ee{roo}   
where $\rho^\mathbf{F}$ is defined as in the equilibrium preparation, 
eq.~(\ref{rb}), and 
$O_j$ and $O_j'$ are operators that only act on the system's
Hilbert space. For applications of this preparation 
class we refer the reader to Ref.~\cite{gsi}.

The state of the open system after the preparation results as the
trace over the environment
of the density matrix of the full system at that time, i.e.   
\be
\rho_S(t) = \Tr_B U(t)\: \rho \: U^\dagger(t),
\ee{rt}
where 
\be
U(t) = \exp \left \{ -\frac{i}{\hbar} H t \right \}
\ee{U}
is the unitary time evolution operator of the full system and $\rho(0) =
\rho$ the density matrix resulting from the preparation.

Requiring an affine time evolution of the reduced density matrix
$\rho_S(t)$ 
means that  
$\rho_S(t)$
is the sum of a term linear in, and one independent of $\rho_S$, i.e.  
\be
\rho_S(t) = T(t)\, \rho_S + I(t),
\ee{T} 
where $T(t)$ is a linear
operator and I(t) is independent of $\rho_S$. 
According to eq.~(\ref{rt}) the
reduced density matrix at time $t$ is a linear image of the initial
full density matrix $\rho$ under the successive action of the unitary time
evolution of the full system and the operation of the trace.  
In order to obtain a map from the initial reduced density matrix
$\rho_S$ to its value at a later time $t$ we introduce the {\it blow-up}
  map 
$R(\rho_S)$ that assigns to each reduced  initial density matrix $\rho_S$
one belonging to the total system: 
\be
\rho = R(\rho_S).
\ee{rRr}
Its particular form  
depends on the initial preparation of the system.
Expressing the initial total density matrix $\rho$ with the help of the
blow-up map by the reduced initial state one obtains the reduced time
evolution of the system:
\be
\rho_S(t) = \Tr_B U(t)\, R(\rho_S)\, U^\dagger(t).
\ee{rSt}
An affine time evolution as described in eq.~(\ref{T}) will result
only from an affine blow-up map.
So one may ask under which conditions the
blow-up map is affine.

Because a density matrix is a positive
normalized operator, we require that the blow-up map
$R$ acts on a convex set of reduced density matrices, i.e. a set that
contains with each pair $\rho_{S1}$ and $\rho_{S2}$ all convex
combinations $\lambda \rho_{S1} + (1-\lambda)\rho_{S2}$ for all $0 < \lambda <1$. We
assume next that the considered preparation provides such a convex set of
reduced density matrices. For any particular preparation one has to
check this property.  We then further
may ask under which conditions $R$ preserves the convexity
condition, put differently under which conditions 
a full density matrix that corresponds to a convex combination
of reduced density matrices again is given by a convex combination. Then
the preparation class     
also forms a convex set. For such a convex blow-up
map one obtains:
\be
R\bm{\bigl (} \lambda \rho_{S,1} + (1-\lambda) \rho_{S,2} \bm{\bigr )} = 
\lambda R(\rho_{S,1}) +(1- \lambda) R(\rho_{S,2}).
\ee{Rc}     
This then implies that the blow-up map is affine:
\be
R(\rho_S) = L \rho_S + \chi,
\ee{Raff}
where $L$ is a linear operator that maps reduced density matrices onto 
full density
matrices and $\chi$ is an operator of the full system. 
A proof of this statement is given in the Appendix~\ref{A2}.

We note that the mixing parameter $\lambda$ has to be identical on the left
and on the right hand side of the eq.~(\ref{Rc}). This is a
consequence of the fact that the trace of $R(\rho_S)$ 
over the Hilbert space of the environment must coincide with the system density
matrix $\rho_S$: 
\be
\Tr_B R(\rho_S) = \rho_S.
\ee{trR}
 
For the factorizing preparation (\ref{rfac}) 
the blow-up map $R$ is always linear. It 
simply acts as the multiplication by the reference environment density
operator $\rho_B$. In the case of a classical system dynamics 
the role of density matrices is taken over by probability densities
defined on the
respective phase space. 
Then, any preparation can be
characterized by a conditional probability density 
$\rho(\mathbf{x}| \mathbf{x}_S)$ for the state 
$\mathbf{x}$ of the total system given the state $\mathbf{x}_S$ of the system
\cite{gth,gtht,ght}.
The corresponding blow-up map $R$ is then given by the multiplication
with this conditional probability and, hence, is always
linear. No such simple construction scheme 
is available in quantum mechanics and 
Pechukas \cite{P} showed that the factorizing preparation is the only
one for which the blow-up map is linear, provided that a sufficient
number 
of pure states of the reduced system are contained in the
preparation class. For the convenience of the reader 
the precise formulation of the theorem and a
proof is given in the Appendix~\ref{A1}.

In the present work, we will consider the influence of the preparation 
of an open quantum system on its 
reduced time evolution by means of the simple example of two
interacting spins. One of those is considered as the
system, the other one plays the role of the environment. The second spin 
is only a very crude 
caricature of a true environment which clearly fails to cause dissipation or
dephasing in the system because of its finiteness. 
Nevertheless, it suffices to illustrate the influence of
the preparation on the subsequent dynamics of the system.     

We assume that the total system starts from an equilibrium
preparation, i.e. 
that its initial state is described by a density matrix
of the form of eq.~(\ref{rb}). 
It
will be  shown that in general this preparation renders the time
evolution of the reduced system 
{\it nonlinear}. 
\section{Two spins}
Both interacting spins 
 $\boldsymbol{\sigma}_1 = (\sigma^x_1,\sigma^y_1,\sigma^z_1)$ 
and $\boldsymbol{\sigma}_2=(\sigma^x_2,\sigma^y_2,\sigma^z_2) $ with Pauli spin
 matrices $\sigma_\alpha^i$, $\alpha=1,2$, $i =x,y,z$, are of total length $s=1/2$. 
The first spin, $\boldsymbol{\sigma}_1$ is considered as the system
and the second one, $\boldsymbol{\sigma}_2$ takes over 
the role of the environment.   
Every density matrix of the total system then assumes the form
\be
\rho = \frac{1}{4} \left ( 1 + \mathbf{S}_1 \cdot \boldsymbol{\sigma}_1 +  
\mathbf{S}_2 \cdot \boldsymbol{\sigma}_2 + \boldsymbol{\sigma}_1 \cdot \mathbf{C}
\cdot \boldsymbol{\sigma}_2 \right ),
\ee{r}
where 
\be
\mathbf{S}_\alpha = \langle \boldsymbol{\sigma}_\alpha \rangle 
\ee{S}
denotes
the Bloch vector of the spin $\alpha =1,2$ and the matrix  
\be
\mathbf{C} = \langle \boldsymbol{\sigma}_1  \boldsymbol{\sigma}_2 \rangle
\ee{C}
denotes the correlation matrix of the two spins. The dot-product denotes the
scalar product in three dimensions, e.g. $ \mathbf{S}_1 \cdot
\boldsymbol{\sigma}_1= S_1^x \sigma_1^x + S_1^y \sigma_1^y + S_1^z \sigma_1^z $.
The reduced density matrix of the system is given by the trace over
the environment (i.e. the second spin), and hence becomes
\be
\rho_S = \Tr_2 \rho = \frac{1}{2} \left ( 1 + \mathbf{S}_1 \cdot
  \boldsymbol{\sigma}_1 \right ).
\ee{rS2}
Here we want to study the opposite direction, that is to go 
from $\rho_S$ to $\rho$.
In particular, we   
look for conditions under which the respective 
blow-up map $R(\rho_S) = \rho$ is convex. We recall that the 
blow-up map is determined by 
the preparation process of the system. 
In the process of a preparation the state of the
system is controlled  by external fields $\mathbf{F}$ that
ideally act only on the system, as given in eq.~(\ref{rb}) for the
equilibrium preparation. For the considered spin one can think
of static magnetic fields. For interacting spins the two Bloch vectors and the
correlation matrix will depend on the applied magnetic field. 

We next investigate the requirement of the convexity of the blow-up map
$R$.
A necessary condition for this property to hold is the
convexity of the domain of definition of $R$. This implies that for
any pair of reduced density matrices $\rho_S^{\mathbf{F}_k}$, $k=1,2$ 
that result from two different
values of the field, all convex linear combinations can be prepared by 
means of another value  $\mathbf{F}_3$ of the field:
\be
\rho_S^{\mathbf{F}_3} = \lambda \rho_S^{\mathbf{F}_1} + (1-\lambda)
\rho_S^{\mathbf{F}_2},
\ee{con}
with $\mathbf{F}_3$ being a uniquely defined function of the fields
$\mathbf{F}_1$, $\mathbf{F}_2$ and $\lambda$. Using the general form of the
reduced density matrix in (\ref{rS2}) one finds a respective equation for
the Bloch vectors, reading
\be
\mathbf{S}_1(\mathbf{F}_3)  = \lambda \mathbf{S}_1(\mathbf{F}_1) +(1-\lambda) 
\mathbf{S}_1(\mathbf{F}_2).
\ee{Sinv}
This means that the Bloch vector of the system must be a  uniquely 
invertible function of the external field. 
For the equilibrium preparation this is the case because then the
derivatives of the Bloch vector components with respect to the field
components coincide with the
correlations of fluctuations of the first spin. These derivatives form
the elements of the susceptibility matrix which is 
an invertible matrix in thermal equilibrium. 
Hence, $\mathbf{S}_1(\mathbf{F})$ has a uniquely defined inverse
$\mathbf{F}(\mathbf{S}_1)$.  
We  investigate the consequences of the convexity condition
(\ref{Rc}) of the blow-up map for the spin system.
Using eqs.~(\ref{con}) and (\ref{r}) 
for the total density matrix of the full system one
finds analogous relations  from eq.~(\ref{Rc})
for both the Bloch vector of the second spin and for the correlation
matrix
with the same field $\mathbf{F}_3$ that results from
eq.~(\ref{Sinv}), i.e.,
\ba
\mathbf{S}_2(\mathbf{F}_3) & = & \lambda \mathbf{S}_2(\mathbf{F}_1) +(1-\lambda) 
\mathbf{S}_2(\mathbf{F}_2), \nonumber \\
\mathbf{C}(\mathbf{F}_3) & = & \lambda \mathbf{C}(\mathbf{F}_1) +(1-\lambda) 
\mathbf{C}(\mathbf{F}_2).
\ea{Cinv}
These are non-trivial conditions which in general will not be satisfied.
Expressing next the fields $\mathbf{F}_i$ by the set of Bloch vectors
$\mathbf{S}^i_1 = \mathbf{S}_1(\mathbf{F}_i) $ that result for the
respective fields we find by use of 
eq.~(\ref{Sinv}) the following relations for the Bloch vectors of the
second spin and the correlation matrix:
\ba
\mathbf{S}_2[\lambda \mathbf{S}^1_1 +(1-\lambda) \mathbf{S}^2_1] & = & \lambda
\mathbf{S}_2[\mathbf{S}^1_1] +(1-\lambda) \mathbf{S}_2[\mathbf{S}^2_1],
\nonumber \\
\mathbf{C}[\lambda \mathbf{S}^1_1 +(1-\lambda) \mathbf{S}^2_1] & = & \lambda
\mathbf{C}[\mathbf{S}^1_1] +(1-\lambda) \mathbf{C}[\mathbf{S}^2_1].
\ea{SCcon}
Hereby  we introduced the notation
$\mathbf{S}_2[\mathbf{S}_1] = \mathbf{S}_2(\mathbf{F})$ and 
$\mathbf{C}[\mathbf{S}_1] = \mathbf{C}(\mathbf{F})$.
From these equations it follows that both $\mathbf{S}_2[\mathbf{S}_1]$
and $\mathbf{C}[\mathbf{S}_1]$ are affine functions, see
Appendix~\ref{A2}. 
Therefore they can be represented as
\ba
\mathbf{S}_2[\mathbf{S}_1]& = & \mathbf{A} \cdot \mathbf{S}_1 +
\mathbf{B}, \nonumber \\
\mathbf{C}[\mathbf{S}_1]& = & \mathbf{D} \cdot \mathbf{S}_1 +
\mathbf{E},
\ea{SCaf}
where $\mathbf{B}$ is a constant vector, $\mathbf{A}$ and $\mathbf{E}$
are constant, second order tensors and $\mathbf{D}$ is a constant third
order tensor. Hence, these quantities must neither depend on 
the applied field $\mathbf{F}$ nor on the Bloch vector $\mathbf{S}_1$.

\section{An explicite illustration}

We consider the equilibrium preparation for the following simple
two-spin Hamiltonian as an example, 
\begin{equation}
\label{H}
H =   
- F_z \sigma^z_1
+ e \sigma_2^z + g \sigma_1^x \sigma_2^x,
\end{equation}
Here we only allow for a field in the $z$-direction. The
two spins interact by their $x$-components. 
We study the equilibrium preparation class at the fixed 
inverse temperature $\beta$ that results if the field $F_z$ 
assumes all possible values, $-\infty < F_z < \infty$: 
\be
\rho^{F_z} = Z^{-1} \exp \left \{ - \beta H \right \}.
\ee{rFz}
Because the Hamiltonian $H$ commutes with the operator $\sigma^z_1 \sigma^z_2$ one can
diagonalize $H$ in the eigenspaces of  $\sigma^z_1 \sigma^z_2$.
This then yields the four eigenvalues $\mathcal{E}_i$ 
\begin{eqnarray}
\nonumber
\mathcal{E}_1 = -\sqrt{(F_z-e)^2+g^2}, & \quad &
\mathcal{E}_2 = \sqrt{(F_z-e)^2+g^2}, \\
\label{Energies}
\mathcal{E}_3 = -\sqrt{(F_z+e)^2+g^2}, & \quad &
\mathcal{E}_4 = \sqrt{(F_z+e)^2+g^2},
\end{eqnarray}
and the corresponding eigenprojection operators $P_i$:
\begin{eqnarray}
P_{i} & = & \frac{1}{4}
        \left(
        1 + \sigma_1^z \sigma_2^z 
        -\frac{F_z-e}{\mathcal{E}_{i}} (\sigma_1^z+\sigma_2^z)
        +\frac{g}{\mathcal{E}_{i}}
                (\sigma_1^x \sigma_2^x-\sigma_1^y \sigma_2^y)
        \right),
\quad i=1,2,
\\
P_{i} & = & \frac{1}{4}
        \left(1 - \sigma_1^z \sigma_2^z 
        -\frac{F_z+e}{\mathcal{E}_{i}} (\sigma_1^z-\sigma_2^z)
        +\frac{g}{\mathcal{E}_{i}}
                (\sigma_1^x \sigma_2^x+\sigma_1^y \sigma_2^y)
        \right),
\quad i=3,4,
\end{eqnarray}
such that $H=\sum_i \mathcal{E}_i P_i$ holds.
The canonical density matrix $\rho^{F_z}$ at the inverse temperature $\beta$ is a
mixture of the pure states $P_i$ with the Boltzmann weights 
$p_i = \exp \left \{ -\beta \mathcal{E}_i \right \} / \left [2 \left ( \cosh (\beta
  \mathcal{E}_1) +\cosh (\beta
  \mathcal{E}_2) \right )\right ]$, i.e.,
\be
\rho^{F_z} = \sum_{i=1}^4 p_i P_i = \frac{1}{4} \left ( 
1 + S_{1z} \sigma^z_1  + S_{2z} \sigma^z_2 + C_{xx} \sigma^x_1 \sigma^x_2  + 
C_{yy} \sigma^y_1 \sigma^y_2  + C_{zz} \sigma^z_1 \sigma^z_2        \right ).
\ee{rfz}
The $x$- and $y$-components of the two Bloch vectors vanish.
The non-vanishing $z$-components read
\ba
S_{1z}& = & \beta \left ( F_z \mathcal{F}_+(\beta \mathcal{E}_1, \beta
  \mathcal{E}_3) - e\mathcal{F}_-(\beta \mathcal{E}_1, \beta
  \mathcal{E}_3)\right ),   \nonumber \\
S_{2z}& = & \beta \left ( F_z \mathcal{F}_-(\beta \mathcal{E}_1, \beta
  \mathcal{E}_3) - e\mathcal{F}_+(\beta \mathcal{E}_1, \beta
  \mathcal{E}_3) \right ),
\ea{S12}
where the auxiliary functions $\mathcal{F}_\pm(x,y)$ are defined by:
\be
\mathcal{F}_\pm (x,y) =  \frac{y \sinh(x) \pm x \sinh(y)  }{x y
  \left (\cosh(x)+\cosh(y) \right ) }.
\ee{Fpm}
Finally, the non-vanishing elements of the correlation matrix $\mathbf{C}$ are
given by:
\ba
C_{xx}& = & -\beta g\mathcal{F}_+(\beta\mathcal{E}_1,\beta\mathcal{E}_3),
\nonumber \\
C_{yy} &= & 
\beta g \mathcal{F}_-(\beta\mathcal{E}_1,\beta\mathcal{E}_3),
\nonumber \\
C_{zz} & = & \frac{\cosh(\beta \mathcal{E}_1) - \cosh(\beta \mathcal{E}_3)} 
{\cosh(\beta \mathcal{E}_1) + \cosh(\beta \mathcal{E}_3)}.
\ea{Cxyz}
\begin{figure}
\includegraphics[width=8cm]{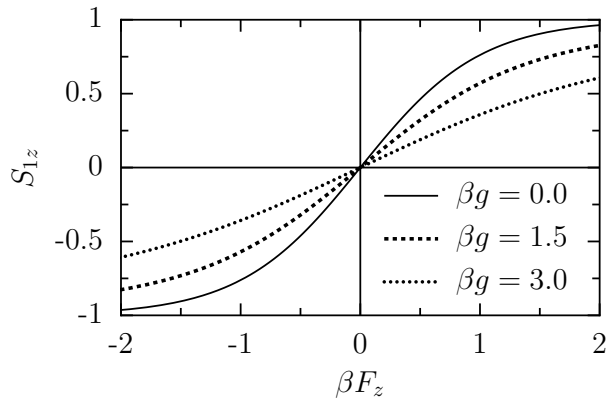}
\caption{The Bloch vector component $S_{z1}$ as a function 
of the field $\beta F_z$  
resulting from eq.~(\ref{S12}) for $\beta e=1$. 
With increasing coupling parameter $\beta g = 0.5, \; 1, \; 1.5$ the
slope at 
$F_z =0$ decreases. Apparently, the functions $S_{z1}(F)$ are
monotonic and hence possess a unique inverse. }
\label{f1}
\end{figure}
As already stated above, in the present case of the equilibrium preparation
the $z$-component of the Bloch vector of the first spin is a
uniquely invertible function of the field magnetic $F_z$. This can be shown by
inspection from eq.~(\ref{Fpm}), see also Fig.~\ref{f1}. 
We note that the  equilibrium preparation contains the pure system
states $\rho_S=\frac{1}{2} \left ( 1 \pm \sigma_z \right )$ asymptotically
in the infinite field limit $F_z \to \pm \infty $. 
If one also took into account
fields that couple to the other spin components $\sigma_x$, $\sigma_y$, the
eigenstates of these components could also be prepared
asymptotically. In the case studied here, however, no other pure states
than the eigenstates of $\sigma_z$ can be prepared. Thus, for this
particular case one of the conditions of the Pechukas theorem to hold are
not met.

\subsection{Testing convexity of the blow-up map}
We now come to the discussion of the eqs.~(\ref{SCaf}) which are necessary
conditions for the blow-up map to be convex. Because in the considered
preparation class only the $z$-component of the field is varied and
because of the symmetries of the considered Hamiltonian these
equations need only be checked as functions of $S_{z1}$ for the Bloch vector
component $S_{z2}$ and the diagonal elements of the correlation
matrix, respectively:

\ba
S_{z2}[S_{z1}] & = & A S_{z1} + B, \nonumber \\
C_{xx}[S_{z1}] & = & D_{xx} S_{z1} + E_{xx}, \nonumber \\
C_{yy}[S_{z1}] & = & D_{yy} S_{z1} + E_{yy}, \nonumber \\
C_{zz}[S_{z1}] & = & D_{zz} S_{z1} + E_{zz}. 
\ea{SCxyz}
If the blow-up
map were convex these equations would have to result from the eqs.~(\ref{S12},
\ref{Cxyz}) by eliminating the 
external field
$F_z$. There is no need to
perform the lengthy calculation to see that the above equations 
hold only if the system-environment interaction is absent, i.e. if the
coupling constant $g$ in the Hamiltonian vanishes. We note that with
$\mathcal{E}_1(-F_z) = \mathcal{E}_3(F_z)$ and the symmetries of the
auxiliary functions 
$\mathcal{F}_\pm(x,y) =\pm \mathcal{F}_\pm(y,x)$, 
$S_{1z}$ becomes an odd function whereas $C_{xx}$ is an even 
function of $F_z$. If eq.~(\ref{SCxyz}) was to hold, $D_{xx}$ would have to
vanish and $C_{xx}$ would have to be a constant. 
This is true if and only if $g=0$.  
\begin{figure}
\includegraphics[width=8cm]{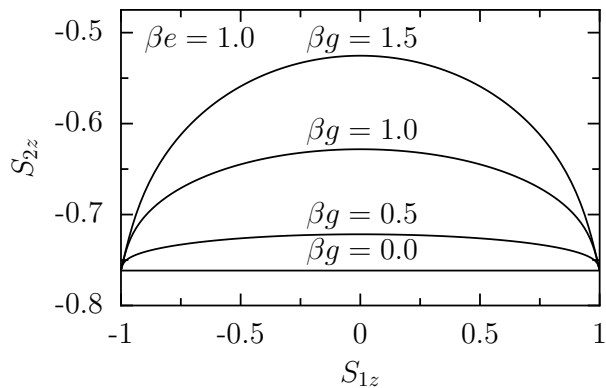}
\caption{The Bloch vector component $S_{z2}$ as a function 
of $S_{z1}$ resulting from eq.~(\ref{S12}) for $\beta e=1$ and different
  values $\beta g = 0, \; 0.5, \; 1, \; 1.5$. Note that
a strictly linear dependence results  only for the case of 
vanishing coupling $g=0$. An approximate linear regime exists in a
  neighborhood of $S_{1z} =0$.}  
\label{f2}
\end{figure}
\begin{figure}
\includegraphics[width=8cm]{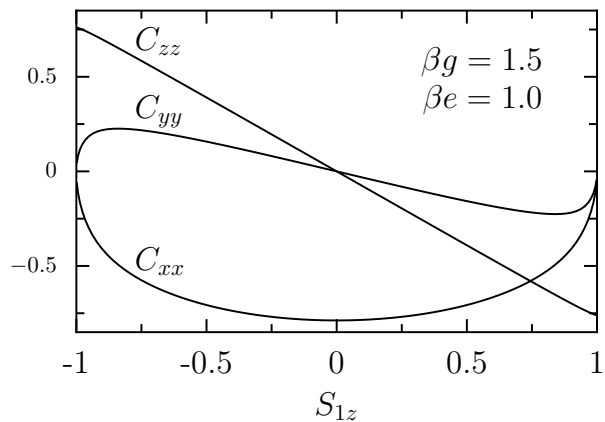}
\caption{The non-vanishing correlation functions $C_{xx},\; C_{yy},
  \;C_{zz}$ as a function of the Bloch vector component $S_z$ for $\beta e
  =1$ and $\beta g =1.5$. The correlation functions $C_{xx}$ and $C_{yy}$
  clearly deviate from straight lines following from
  eq.~(\ref{SCxyz}).  For the correlation functions also an 
approximately linear regime exists in a
  neighborhood of $S_{1z}=0$.  }
\label{f3}
\end{figure}
Figs.~\ref{f2} and \ref{f3} depict the dependences of $S_{z2}$ and  
of the correlation functions, respectively, on  $S_{z1}$ for
finite coupling strengths.  In all cases, except for $g=0$, 
and for the correlation $C_{zz}$,
the deviations from linearity are strikingly obvious. We note,
however, that for small values of the Bloch vector
component $S_{1z}$ the component 
$S_{2z}$ and correlation $C_{xx}$ are almost constant and the other 
correlations 
$C_{zz}$ and $C_{yy}$ are linear. This behavior is in accordance 
with a convex blow-up map.     
Below we will come back to the blow-up in the linear response regime when
the external fields are small.

\section{Discussion, implications and conclusions}
We have illustrated Pechukas' verdict on the linear time evolution of open
quantum systems \cite{P}  by a simple example. 
Moreover, we have  
demonstrated that the theorem holds for affine time evolutions: 
If two complete sets of pure system states can be prepared this more general
class of evolution implies 
a factorizing preparation of the total initial density
matrix where the environment density matrix must be independent
of the system density matrix.
Actually, Pechukas' original proof \cite{P} applies as well in the affine
case. Nowhere in the proof he made explicit use of the homogeneity
condition, i.e. 
$R(\lambda \rho_S )= \lambda R(\rho_S)$, $\lambda$ real,  that would
render an affine $R$ a linear map.

At first, this may only seem a modest generalization of the original
conclusion but it sheds some light on the crucial role of the 
inhomogeneity term of
(generalized) master equations which appears when the initial density
matrix does not factorize. The present analysis excludes that this
term is {\it independent} of the initial reduced density matrix and
actually is not merely an inhomogeneity of the otherwise linear master
equation but must depend on the reduced density matrix in a {\it nonlinear
way}. In those cases, when a Markovian dynamics is approached for long
times this nonlinear term must vanish for sufficiently large times. It 
will do so, however, in a characteristic manner 
that depends on the particular initial reduced
density matrix.  

In the present example only one set of pure system states, the
eigenstates of $\sigma_z$,  are preparable and yet the affinity of the
blow-up map implies
a factorizing preparation. We note that in general the condition  on
the number of preparable pure states of
the Pechukas theorem cannot be relaxed. A counter example 
is provided by the work of
Karrlein and Grabert \cite{KG} who considered a harmonic 
oscillator coupled to a bath
of harmonic oscillators with a non-factorizing thermal preparation. 
This preparation allows for pure
position states and still leads to a linear master equation. 

Another example of a non-factorizing
preparation that leads to a linear master equation results from the
following preparation procedure: (i) start with a factorizing density
matrix $\rho_{S0}\: \rho_{B0}$ 
at a time $t=-t_0$, $t_0 >0$, (ii) turn on the interaction between
the system and the bath, and (iii) use the density matrix that has evolved
at $t=0$ as the result of the preparation. We term this preparation the 
{\it factorize-and-wait} preparation. The blow-up map $R(\rho_S)$ then assumes
the form:
\be
R(\rho_S) = e^{-\frac{i}{\hbar}H t_0} G_{t_0}^{-1}( \rho_{S}) \rho_{B0}  
e^{\frac{i}{\hbar}H t_0}, 
\ee{but0}
where $G_t$ is the linear propagator of the reduced density matrix for the
factorizing preparation, i.e.,
\be
\rho_S(t)=G_t(\rho_S(0)) = \tr_B  e^{-\frac{i}{\hbar}H t_0}
\rho_{S}(0) 
\rho_{B0}  
e^{\frac{i}{\hbar}H t_0}. 
\ee{Gt}
Here, the inverse of the propagator $G_t$ is needed in order to infer the
proper system part of the factorizing density at $t=-t_0$ from the
density matrix $\rho_S$ that is to be prepared at $t=0$. In view of
possible fast relaxation processes and the build-up of system-bath 
correlations \cite{Be,HR,O}
after the interaction has been switched on, this inverse propagator will not be
defined on the total set of possible density matrices \cite{HTGT}. 
Still, it is a linear operator and thus the blow-up map (\ref{but0}) also
is linear. 
It is obvious, however, that no pure states of the system can be
prepared at time $t=0$ in
this way, i.e., that the factorize-and-wait preparation does not
provide pure states. Therefore, the conditions for the Pechukas theorem
are not met and the theorem is thus not in conflict with the linearity of
the blow-up map of the entangled factorize-and-wait preparation.    

Finally, we discuss the equilibrium preparation in the limit of weak
external forces. 
In the region close to thermal equilibrium 
Mori's generalized quantum
Langevin equations \cite{Mori} provide a proper description of the
time evolution of the set of system operators that couple to the
external fields. In this case, 
the preparable density matrices of the total system
follow from eq.~(\ref{rb}) in linear approximation in the external
fields $F_j$ and hence assume the linear response form \cite{Kubo}:
\be
\rho^{\bF}= \rho^\bN + \beta \sum_i \int_0^1 d x \:(\rho^\bN)^{1-x} 
\left ( X_i -  \langle X_i \rangle_0 \right ) (\rho^\bN)^x F_i,
\ee{rtlr} 
where $\rho^\bN$ denotes the equilibrium density matrix of the total
system in the absence
of external fields and $\langle X \rangle_0$ the average of the
operator $X$ with respect to the density matrix $\rho^\bN$.  
The corresponding density matrix of the reduced
system $\rho_S^\bF = \Tr_B\rho^\bF$ then becomes
\be
\rho_S^\bF = \rho_S^\bN +   \beta \sum_i \int_0^1 d x \:\Tr_B  
(\rho^\bN)^{1-x} 
\left ( X_i -  \langle X_i \rangle_0 \right ) (\rho^\bN)^x F_i.
\ee{rrlr}
We recall that the external fields act on the system, i.e. that the
conjugate operators $X_i$ are system operators. The expectation
values of these operators 
with respect to the density matrix $\rho_S^\bF$ are linear
functions of the external fields by construction and can be written as:
\be
\langle X_j - \langle X_j \rangle_0 \rangle \equiv \Tr_S \left (X_j -
  \langle X_j \rangle_0 \right ) \rho_S^\bF = \sum_i \chi_{ji} F_i,
\ee{XF}
where the response matrix $\chi_{ij}$ is obtained by inserting
eq.~(\ref{rtlr}) into the middle term of eq.~(\ref{XF}).
\be
\chi_{ij} = \beta \int_0^1 dx\: \Tr \left (X_i-\langle X_i \rangle \right )
(\rho^\bN)^{ 1-x}  \left (X_j-\langle X_j \rangle_0 \right ) (\rho^\bN)^x.
\ee{chi}
Because $\chi_{ij}$ is an invertible matrix  we 
may solve eq.~(\ref{XF}) for the external
fields $F_i$
\be
F_i = \sum_j \chi^{-1}_{\; \;\;\;ij} 
\langle X_j - \langle X_j \rangle_0 \rangle.
\ee{FX}
In this way the external fields are expressed in terms of the linear 
functional $\langle X_j -  \langle X_j \rangle_0  \rangle$
of the system density matrix $\rho_S^{\bF}$. Using this relation in
eq.~(\ref{rtlr}) we find for the blow-up map the affine form
$R(\rho^\bF_S) = L \rho^\bF_S + \chi$ where
\ba
L \rho^\bF_S &=& \beta \sum_{i,j} \int_0^1 dx\; (\rho^\bN)^{1-x} \left (
  X_i -\langle X_i \rangle_0 \right ) (\rho^\bN)^x \chi^{-1}_{\;\;\;\;ij} 
\Tr_S\left ( X_j - \langle X_j \rangle_0 \right ) \rho^\bF_S, \nonumber \\
\chi &=& \rho^\bN .
\ea{Rlr} 
With eq.~(\ref{rt}) one obtains for the time evolution of the reduced
density matrix 
\be
\rho^\bF_S(t) = \Tr_B U(t) L(\rho^\bF_S) U^\dagger(t) + \Tr_B \rho^\bN
\ee{te}
where we used that the total density matric $\rho^\bN$ is invariant
under the full time evolution. 
Actually this is a linear equation in the reduced density matrix
$\rho_S$. In order to see this one
puts $t=0$
in eq.~(\ref{te}), 
to express the 
trace of $\rho^\bN$ over the
environment 
in terms of the reduced density
matrix $\rho_S$: $ \Tr_B \rho^\bN = \rho_S - \Tr_B L(\rho_S)$. 
Hence, for the Mori preparation 
the time evolution of the reduced density matrix is linear. This is
also in agreement with the findings for the above discussed model, see
Figs.~\ref{f2} and \ref{f3}.

\appendix  
\section{Convex maps are affine}\label{A2}
We prove that a differentiable convex map $M$ from a Banach space
$B_1$ into a Banach space $B_2$ is also affine. 
The derivative of $M(x)$ at $x \in B_1$ is defined as the linear map 
$DM(x): B_1 \to
B_2$ that is tangential to $M$ at $x$. For further mathematical
details see e.g. Ref.~\cite{AMR}:
\be
\lim_{||h||_1\to 0} \frac{||M(x+h)-M(x)-D\!M(x)h||_2}{||h||_1} = 0,
\ee{diff} 
where $||\cdot||_i$, $i=1,2$ denotes the norm in the respective Banach
space. The convexity of $M$ then requires that its domain of
definition ${\mathcal D}(M)$ is convex, i.e. that with each pair
$x,\;y$ of elements of  ${\mathcal D}(M)$ also all convex linear 
combinations, $\lambda x +(1-\lambda)
y$, with $0 <\lambda <1$, belong to  ${\mathcal D}(M)$. Moreover, this
property 
is conserved under the convex map $M$:
\be
M\bigl (\lambda x +(1-\lambda) y \bigr ) = \lambda M(x) +(1-\lambda) M(y).
\ee{cM}
Taking the derivative with respect to $\lambda$ one finds from (\ref{cM}):
\be
D\!M\bigl (\lambda x + (1-\lambda) y \bigr )(x-y) = M(x) -M(y).
\ee{DM}
For $\lambda = 0$ one obtains
\be
M(x) = D\!M(y)x +M(y) - D\!M(y)y.
\ee{Mx}
The first term on the right hand side is linear with respect to $x$
and the second and third term are independent of $x$ and hence M is affine. 

\section{Pechuka's theorem}\label{A1}
Pechukas proved in Ref.~\cite{P} that a preparation is factorizing
if the following conditions are satisfied: (i) the corresponding blow-up map of
the preparation is convex; (ii) two different complete sets of pure
system states can be prepared in the considered preparation class.

The proof consists of two steps. The first step is a consequence of
the positivity of the blow-up map $R(\rho_S)$ and 
makes use of the possibility to
prepare pure states of the system. Assume $\{|\psi \rangle \}$ is a 
pure state of the system that can be prepared. In the first step of
the proof it is shown that the density matrix $\rho$ of the full system
has the form
\be
\rho = R(|\psi \rangle \langle \psi |) = |\psi \rangle \langle \psi | \chi
\ee{Rps}
where the bath-density matrix $\chi$ in general depends on the system
state $|\psi \rangle$. 

For the proof we note that any density matrix of the total sytem can
be represented as a weighted sum of products of pure system
density matrices $|\psi_i \rangle \langle \psi_i |  $   
and bath density matrices $\chi_i$,
\be
\rho= \sum_{i} c_i | \psi_i \rangle \langle \psi_i | \chi_i,
\ee{rc} 
where the states
$|\psi_i \rangle$ are normalized and orthogonal on each other:
$\langle\psi_i | \psi_j \rangle = \delta_{i,j}$, and the coefficients
$c_i$ are not negative and add up to unity: $\sum_i c_i =1$. Taking
the trace of $R(|\psi \rangle \langle \psi |)$ over the bath  
one recovers the pure state $|\psi \rangle \langle \psi |$ and on the
other hand, using $\Tr_B \chi_i =1 $ one finds from eq. (\ref{rc})  :
\be
|\psi \rangle \langle \psi | = \sum_i c_i |\psi_i\rangle \langle \psi_i |.
\ee{psps}
Hence, all but one coefficients $c_i$ vanish and eq.~(\ref{Rps}) holds.

Up to this point we have not made use of the convexity of the blow-up
map and therefore the reduced bath density matrix $\chi$ may still
depend on the system state $|\psi \rangle$. 

In the second step of the proof we first take two pairs of orthonormal
system states, $|\psi_1 \rangle $, $|\psi_2 \rangle $ and $|\varphi_1 \rangle
$, $|\varphi_2 \rangle $
that span the same two-dimensional subspace:
\be
|\psi_1\rangle \langle \psi_1 | +|\psi_2\rangle \langle \psi_2 | = 
|\varphi_1\rangle \langle \varphi_1 | +|\varphi_2\rangle \langle \varphi_2 | \equiv P,
\ee{sub}
where $P$ is the projection operator on this subspace.
The absolute values of the mutual scalar products then are determined 
by an angle
$\alpha$:
\ba
|\left (\psi_1,\varphi_1 \right )|^2 = \cos^2 \alpha, \quad 
|\left (\psi_1,\varphi_2\right )|^2 =
    \sin^2 \alpha, \nonumber \\
|\left (\psi_2,\varphi_1 \right)|^2 = \sin^2 \alpha, \quad 
|\left (\psi_2,\varphi_2\right )|^2 =
    \cos^2 \alpha.
\ea{alpha}
We assume that the two pairs of states $\psi_i$ and $\varphi_i$ 
can be prepared. According to the first
step of the proof the full density matrix that corresponds to either
of the pure states is a product of the pure state and a density matrix
of the bath:
\be
R(|\psi_i\rangle \langle \psi_i |) = |\psi_i\rangle \langle \psi_i |
\chi(\psi_i), \quad 
R(|\varphi_i\rangle \langle \varphi_i |) = 
|\varphi_i\rangle \langle \varphi_i |
\chi(\varphi_i), \quad i =1,\; 2,
\ee{ps}  
where $\chi(\psi_i)$ and  $\chi(\varphi_i)$ are bath density
matrices. 
We shall show
that they all are the same. For this purpose we make use of the
convexity of the blow up map and consider its action on the
density matrix that is proportional to the projection $P$ onto the
two dimensional subspace, $\rho_S = \frac{1}{2} P$. 
The action of the blow up map can be represented in
either of two ways by convex combinations of pure states, see eq.~(\ref{sub}):
\be
 |\psi_1\rangle \langle \psi_1 | \chi(\psi_1) +  
|\psi_2\rangle \langle \psi_2 |
 \chi(\psi_2)
= |\varphi_1\rangle \langle \varphi_1 | \chi(\varphi_1) +  
|\varphi_2\rangle \langle \varphi_2 |
\chi(\varphi_2). 
\ee{Rp}
Calculating the matrix elements with the four pure states $\psi_i$,
$\varphi_i$, one obtains the following equations relating the reference
density matrices $\chi(\psi_i)$ and $\chi(\varphi_i)$:
\be
\left (
\begin{array}{c}
\chi(\psi_1) \\
\chi(\psi_2)
\end{array} 
\ \right ) =
\left (
\begin{array}{cc}
|\cos \alpha|^2 & |\sin \alpha |^2 \\
|\sin \alpha|^2 & |\cos \alpha |^2
\end{array}
\right )
\left (
\begin{array}{c}
\chi(\varphi_1) \\
\chi(\varphi_2)
\end{array}
\right ),
\ee{cc1}
and   
\be
\left (
\begin{array}{c}
\chi(\varphi_1) \\
\chi(\varphi_2)
\end{array} 
\ \right ) =
\left (
\begin{array}{cc}
|\cos \alpha|^2 & |\sin \alpha |^2 \\
|\sin \alpha|^2 & |\cos \alpha |^2
\end{array}
\right )
\left (
\begin{array}{c}
\chi(\psi_1) \\
\chi(\psi_2)
\end{array}
\right ).
\ee{cc2}
We may exclude the trivial cases when the pairs 
$\psi_i$ and $\varphi_i$ coincide   
and either $\cos^2 \alpha =1$ or $\sin^2 \alpha =1$. In all other cases,
the four equations only have a solution with $\chi(\psi_1) = \chi(\psi_2)
= \chi(\varphi_1) =\chi(\varphi_2)$. Hence, the bath density matrix is 
identical
for all preparable system density matrices in the considered subspace
such that a factorizable state of the total system results on this subspace. 

In order to apply this argument to higher dimensional system
Hilbert spaces it must be possible to prepare sufficiently many pure
system states. Starting as above with a two dimensional subspace that
can be spanned by two pairs of preparable bases $\{ \psi_i \}$, $\{
\varphi_i \}$, $i= 1,2$ one next considers the
subspace spanned by either $|\psi_1 \rangle $ and $|\psi_3 \rangle $ or
by  $|\varphi_1 \rangle $ and $|\varphi_3 \rangle $ and finds $\chi(\psi_1) =
\chi(\psi_3)$ for the reference bath operators in the density matrix of
the full system. That means that two
different sets of pure states spanning the system's Hilbert space 
must be preparable.

\acknowledgments
The authors thank Gert-Ludwig Ingold and Juan Parrondo   
for valuable discussions and hints. This work was supported by the 
Research network ``Quanteninformation entlang der A8'' and by the
Sonderforschungsbereich 631 of the Deutsche Forschungsgemeinschaft.

\end{document}